\documentclass[runningheads, envcountsame, a4paper]{llncs}
\spnewtheorem{algorithm}[theorem]{Algorithm}{\bfseries}{\rmfamily}
\spnewtheorem{reduct}[theorem]{Reduction}{\bfseries}{\itshape}
\usepackage{graphicx}
\usepackage{amssymb}
\begin{document}
\title{Graph-Theoretic Partitioning of RNAs and Classification of Pseudoknots-II}
%
%
\author{Louis Petingi\inst{1,2}\orcidID{0000-0002-9421-9665}}
%
\authorrunning{L. Petingi}
%
\institute {Computer Science Department, College of Staten Island  (CUNY)
                 \and City University of New York Graduate Center \\
                 \email {louis.petingi@csi.cuny.edu}}
\maketitle              
\begin{abstract} 
Dual graphs have been applied to model RNA secondary structures with pseudoknots, or intertwined base pairs. In previous works, a linear-time algorithm was introduced to partition dual graphs into maximally connected components called blocks and determine whether each block contains a pseudoknot or not. As pseudoknots can not be contained into two different blocks, this characterization allow us to efficiently isolate smaller RNA fragments and classify them as pseudoknotted or pseudoknot-free regions, while keeping these sub-structures intact. Moreover we have extended the partitioning algorithm by classifying a pseudoknot as either recursive or non-recursive in order to continue with our research in the development of a library of building blocks for RNA design by fragment assembly. In this paper we present a methodology that uses our previous results and  classify pseudoknots into the classical H,K,L, and M types, based upon a novel  representation of RNA secondary structures as dual directed graphs (i.e., digraphs). This classification would help the systematic analysis of RNA structure and prediction as for example the implementation of more accurate folding algorithms.

\end{abstract}
%
%
%
\section{Introduction}
%
%

%
%

Let  $G=(V,E)$ be undirected graph composed of  by a finite set of {\it vertices} $V$ and a set $E$ of unordered pairs $e= (v_1,v_2)$  of vertices.  called {\it edges}, where each edge represents a relation between two vertices. 

In 2003,  Gan et. al \cite{gan03} introduced dual graphs to model RNA secondary structures (2D). The 2D elements of RNA molecules consist of double-stranded (stem) regions defined by base pairing such as Adenine-Uracil, \\ Guanine-Cytosine, Guanine-Uracil, and single stranded loops;  stems  and loops are mapped to the vertices and edges of the corresponding dual graph, respectively (later we present an alternative definition of dual graphs). 

Dual graphs can represent complex RNA structures called pseudoknots ({\it PKs}),   which result when two base-paired regions intertwined. Pseudoknots have been associated with a diverse range of important RNA activities as for example in viral gene expression and genome replication (e.g., hepatitis  C, and SARS-CoV viruses). Even though emphasis  has been recently placed on viral translational initiation and elongation, the broader roles of pseudoknots are well-documented \cite {BPG07,PP14}. 

Frameshifting is promoted by an mRNA (i.e., messenger RNA). In programmed ribosomal
frameshifting,  the ribosome is forced to shift one
nucleotide backwards into an overlapping reading frame and to
translate an entirely new sequence of amino acids. This process is essential in the replication of viral pathogens. The signal composed of two essential elements: a heptanucleotide
‘slippery’ sequence
and an adjacent mRNA secondary structure, predominately a pseudoknot.
 
In \cite {Pet1,PetSch} a linear-time partitioning algorithm was introduced based on the dual graph representation of  RNA 2D.  This algorithm partitions a dual graph into connected components called {\it blocks} and then determines whether each block contains a pseudoknot or is a regular region. Thus our procedure provides a systematic approach to partition an RNA 2D, into smaller classified regions, while providing a topological perspective for the analysis of RNAs.

In \cite {PetSch2} Pseudoknots were classified into two main groups: {\it recursive} and {\it non-recursive} pseudoknot. The former is distinguished from the latter because it contains an internal pseudoknotted or regular region that does not intertwine with external stems within the PK. In addition, if the PK is recursive, the partitioning algorithm uniquely identifies each recursive region. 

In the classical literature, pseudoknots have been classified and predicted by folding algorithms into four types: H,K,L, and  M~\cite {KHSQ16}.  Even though each type is defined in terms of how few stems intertwined, pseudoknots can be complex structures, recursive, and be comprised of several stems. In this paper we show based on dual directed graphs, each PK type can be identified through a series of reductions to an unique representative. This methodology will allow us to systematically analyze thousands of motifs and develop more precise RNA folding algorithms. 

The stimulatory nature of PKs in viral replication, and specifically in \\ frameshifting, has been widely studied (see for example [2,4,5,7]).  In a recent publication (2021), Bhatt et al. [2] presented a detailed study of programmed ribosomal frameshifting in translation of the SARS-CoV-2 virus. Also, in [5], structural properties of PKs viral replication in are discussed. Interesting enough evidences show that the simplest H-type pseudoknot (or related structures, see Section \ref{S3}) are predominantly present in eukaryotic families of viral mRNAs. That is not the case in other secondary structures, as pseudoknots could be composed of several intertwining stems (see Section \ref{S1}, Fig. \ref{fig:rf00094}). Related questions follow from this observation;

\begin {enumerate}
\item Are there families of viruses or retroviruses in which other types of PKs (i.e., K, L, M) stimulate frameshifting.
\item Are there structural differences in pseudoknots comprising the FSEs in eukaryotic cells versus prokaryotic cells. 
\item Are there viral mRNAs in which PKs are not present.

\end{enumerate}

Our methodology will be able to shed some light on these outstanding questions. 

In addition, our algorithms allow the study of biological functionality pertaining PKs in general RNA structure as, for  example, to better understand catalytic steps of protein synthesis,  plant viral RNAs with tRNA-like structures, and RNA splicing, among other functions. 


In the next section, we present background material and definitions relevant to this paper, and we review the partitioning algorithm introduced in \cite {Pet1,PetSch}, as well as its applications, as for example the development of a library of building blocks for RNA design by fragment assembly \cite{JBPT18,jain19}; we also review modified partitioning algorithm can detect recursive PKs and its recursive regions.  In section~\ref{S3}, we show how the aforementioned PK types are identified from dual digraphs. In Section~\ref{S4} we summarize the findings. An Appendix section includes computational tests performed by the modified algorithm, on some RNA's motifs.
\section{Background}
\label{S1} \vspace{-4pt}
\subsection {Biological and Topological Definitions}

In 2003, Gan et. al \cite{gan03} introduced {\it tree} and {\it dual} graph-theoretic representations of RNA 2D motifs in a framework called RAG (RNA-As-Graphs)~\cite {fera04,gan04,izzo11,Kim13-1}. Unlike tree graphs, dual graphs detect pseudoknots, thus all the results in this paper are based on the dual graph representation of RNAs 2D.

Topological properties of dual graphs, suggests an alternative  way to look at the  problem of detection and classification of PKs and of general RNAs.  As base pairing in PKs is not well-nested, making the presence of PKs in RNA sequences more difficult to predict by the more classical dynamic programming~\cite{Dirks03} and context-free grammars standard methods \cite{Condon04}.

Following some of the definitions from (Kravchenko, 2009~\cite{Kra09}), we define our biological variables as follows.

\begin {definition} \label{def1} General terms:

\begin {enumerate}
\item [a.] {\it RNA primary sequence}: a sequence of linearly ordered bases $ x_1, x_2, \ldots, x_r$, where $x_i \in \{ A, U,$ $ C, G\}$.
\item [b.] {\it canonical base pair}: a base pair $(x_i, x_j) \in \{(A,U), (U,A), (C,G), (G,C),$ 
                 $(G,U), (U,G)\}$.
\item [c.] {\it RNA secondary structure without pseudoknot - or regular structure, encapsulated in the region $(i_0, \dots, k_0)$}: an RNA 2D structure  in which no two base pairs $(x_i, x_j), (x_l, x_m)$, satisfy $i_0 \leq i < l < j < m \leq m_0$ (i.e., no two base pairs intertwined).
\item [d.] {\it a base pair stem}: a tuple $(x_i, x_{i+1},\ldots,x_{i+r}, x_{j-r},\ldots, x_{j-1}, x_j)$ in which $(x_i, x_j),$ $ (x_{i+1}, x_{j-1}), \ldots, (x_{i+r}, x_{j-r})$ form base pairs.
\item [e.]  {\it segment region}: is a tuple $(x_i, x_{i+1},\ldots,x_{i+r})$  in which $(x_i, x_j)$ is not a base pair whenever $j - i \ge 1$.
\item [f.] {\it a pseudoknot encapsulated in the region} $(i_0, \dots, k_0)$:  if $\exists l,m,  (i_0 < l < m < k_0)$ such that $ (x_{i_0}, x_m)$ and $ (x_l, x_{k_0})$ are base pairs (i.e., at least two base pairs intertwined).
\end {enumerate}
\end{definition}

\begin{figure}
\centering
\includegraphics[scale=.41]{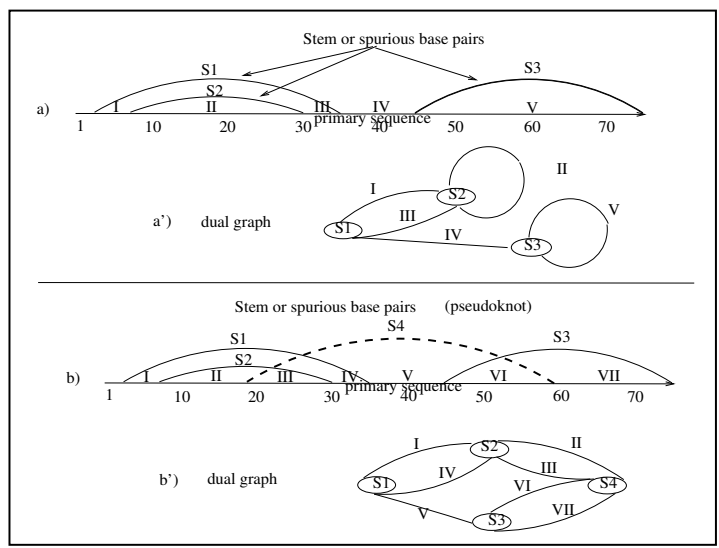}
\caption{\it Graphical and dual graph representations of an RNA 2D structure.
(a) graphical representation of a pseudoknot-free RNA primary sequence and embedded stems or base pairs; (a$^\prime$) corresponding dual graph representation.  (b) graphical representation of a pseudoknotted RNA 2D structure; (b$^\prime$) corresponding dual graph. This figure was originally depicted in \cite{Pet1}.} \label{fig:eulerian}
\end{figure}

A dual graph can be easily derived from the graphical representation of an RNA 2D structure: each stem is modeled by a vertex of the dual graph, and following the primary sequence in linear order (i.e., from left to right), a segment between stems $S_i$ and $S_j$ is represented by an edge $(S_i, S_j)$ in the dual graph (see Fig.~\ref{fig:eulerian}). 

In the next section we present our partitioning approach as of a dual graph $G$, into subgraphs $G' \subseteq G$, called blocks.  
 
\subsection{Graph Partitioning Algorithm}
\label {S2.1}
\begin{figure}[bth]
\begin{center}
\includegraphics[scale=.33]{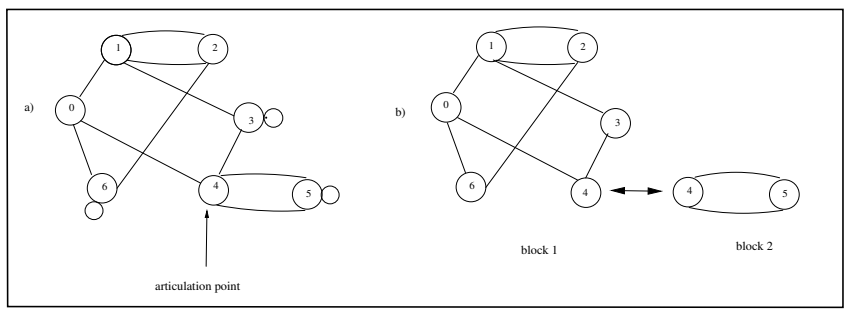}
\end{center}
\caption{\it Identification of a) articulation points and b)  partitioning of a dual graph.} \label{fig:PDB01069}
\end{figure}

The graph-theoretic partitioning algorithm is based on identifying {\it articulation points} of the dual graph representation of an RNA 2D. An articulation point is a vertex of a graph whose deletion disconnects a graph or an isolated vertex remains. 
Articulation points allow us to identify  blocks (see Fig.~\ref{fig:PDB01069}); since  a block is a maximally non-separable component,  a pseudoknot cannot be then contained in two different blocks. Thus identification of these block components allows us to isolate pseudoknots (as well as pseudoknot-free blocks), without breaking their structural properties.

An algorithm for identifying (bi-connected) block components in a graph was introduced by John Hopcroft and Robert Tarjan (1973, \cite {HT1}), and runs in linear computational time.

A {\it hairpin} loop occurs when two regions of the same strand, usually complementary in nucleotide sequence when read in opposite directions, base-pair to form a double helix that ends in an unpaired loop. A self-loop in the dual graph, i.e., an edge having the same vertex as the end-points, represents a hairpin, and as it does not connect two different vertices (i.e., stems), it is formally deleted from the dual graph.

 From Definition~\ref{def1}-c, an RNA 2D structure is a regular-region (pseudoknot-free) and encapsulated in a region $(i_0, \dots, k_0)$, if no two base pairs $(x_i, x_j), (x_l, x_m)$, satisfy $i < l < j < m$, $i_0 \leq i,j,l,m \leq m_0$, otherwise the region is a pseudoknot; this definition yields the following main result.

\begin{corollary}\label{conclude}\cite {Pet1,PetSch}  Given a dual graph representation of  RNA 2D structure, a block represents a pseudoknot if and only if the block has a vertex of degree (Definition \ref{def1}-f) at least $3$ where the degree of a vertex $u$ is the number of edges incident at $u$.
\end{corollary}
Corollary~\ref{conclude} yields the following algorithm,

\setcounter{theorem}{0}
\begin{algorithm} \label{algo:partitioning}{\bf Partitioning}
\begin {itemize}
\item [1.] {\em Partition the dual graph into blocks by application of Hopcroft and Tarjan's algorithm.}
\item [2.] {\em Analyze each block to determine whether contains a vertex of degree at least 3. If that is the case then the block contains a pseudoknot, according to Corollary~\ref{conclude}. If not then the block represents a pseudoknot-free structure.}
\end{itemize}
\end{algorithm}
\setcounter{theorem}{2}

Consider as an example the dual graph shown in  Figure~\ref{fig:PDB01069}. This graph is decomposed into 2 blocks. According to Corollary~\ref{conclude}, block 1 is a pseudoknot as it has a vertex of degree at least 3, while block 2, a cycle, corresponds to a regular region.

 Our partitioning algorithm was applied recently \cite{JBPT18} to analyze 
 the modular units of RNAs for a representative database of experimentally 
 determined RNA structures and to develop a library of building blocks for 
 RNA design by fragment assembly, as done recently for tree graphs, along 
 with supporting chemical mapping experiments \cite{jain18}. Among the 22 
 frequently occurring motifs we found for known RNAs up to 9 vertices, 15 
 contain pseudoknots \cite{JBPT18}. Thus, further classification of the 
 pseudoknotted RNAs could help in cataloging and applications to RNA 
 design. Another application of the partitioning algorithm to small and 
 large units of ribosomal RNAs of various prokaryotic and eukaryotic 
 organisms helped identify common subgraphs and ancestry relationships \cite{JBPT18}.

In the next section we extend our algorithm to classify PKs as either recursive or non-recursive; the algorithm can also identify each recursive region.

\subsection {Classification of pseudoknots as either recursive or non-recursive and identification of each recursive region}
\label{S2.2}
\begin{figure}[bth]
\begin{center}
\includegraphics [scale=0.43]{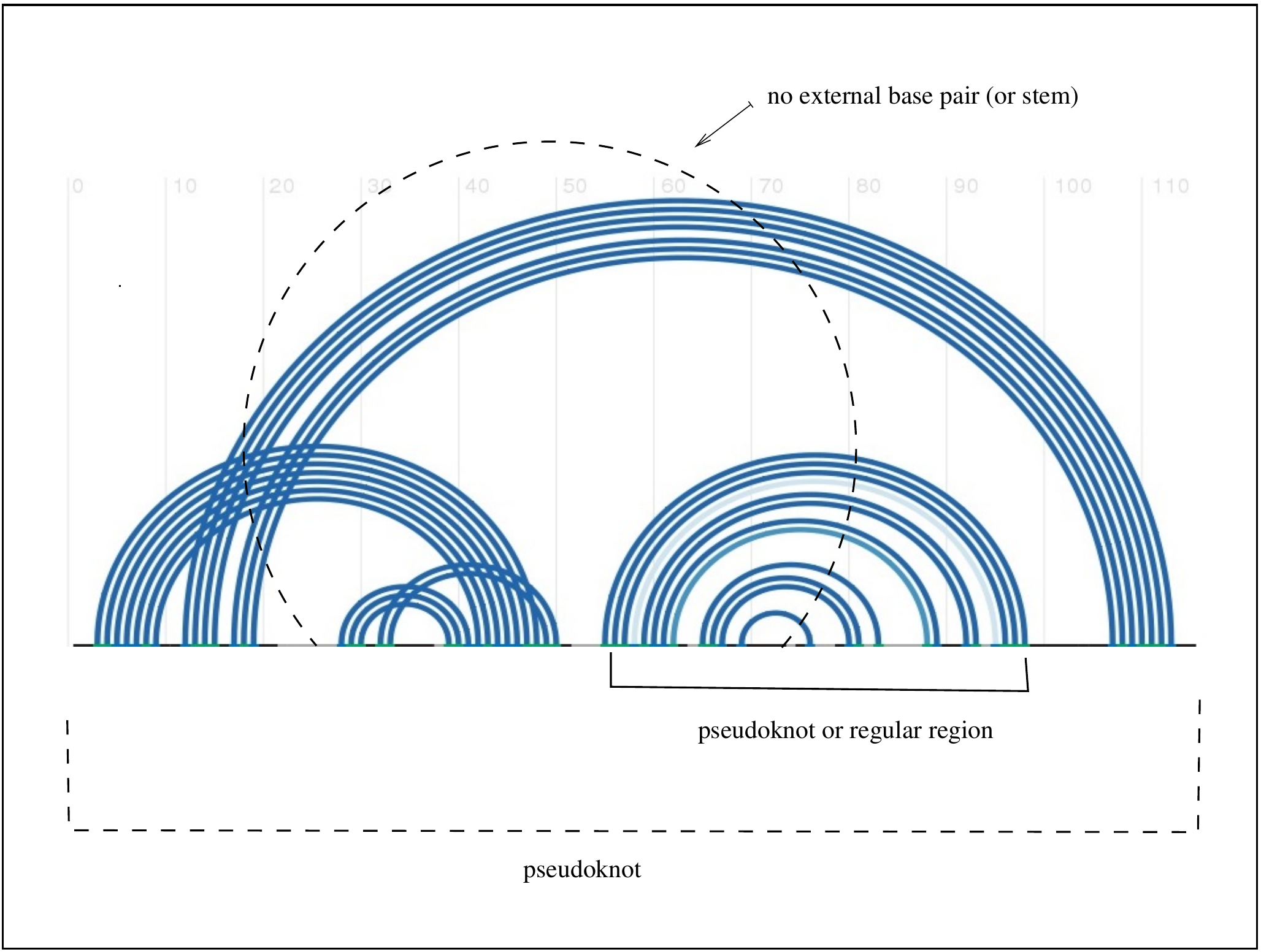}
\end{center}
\caption{\it Recursive pseudoknot.} \label{fig:frecursive}
\end{figure}

The RNA 2D dual graph and graphical representations depicted in this section are based upon New York University's {\it RAG}-database \cite{izzo11}, and {\it R-Chie} visualization software~\cite{ERna}, respectively.

The  definition of  a recursive pseudoknot follows the one of Wong et al. \cite{wong}. A  recursive pseudoknot is a pseudoknot $M_{i,j}$  in a region $[i,j]$ that contains a pseudoknotted or regular region $ M_{k,l}$, $i <  k < l < j$, and there does not exist a base pair $(x_c,x_d)$, such that $x_d$ is a base of $M_{k,l}$, and $x_c$ is a base of $M_{i,j}$ external to  $M_{k,l}$ (see Fig. \ref{fig:frecursive}). 

\begin{figure}[bth]
\begin{center}
\includegraphics [scale=0.32]{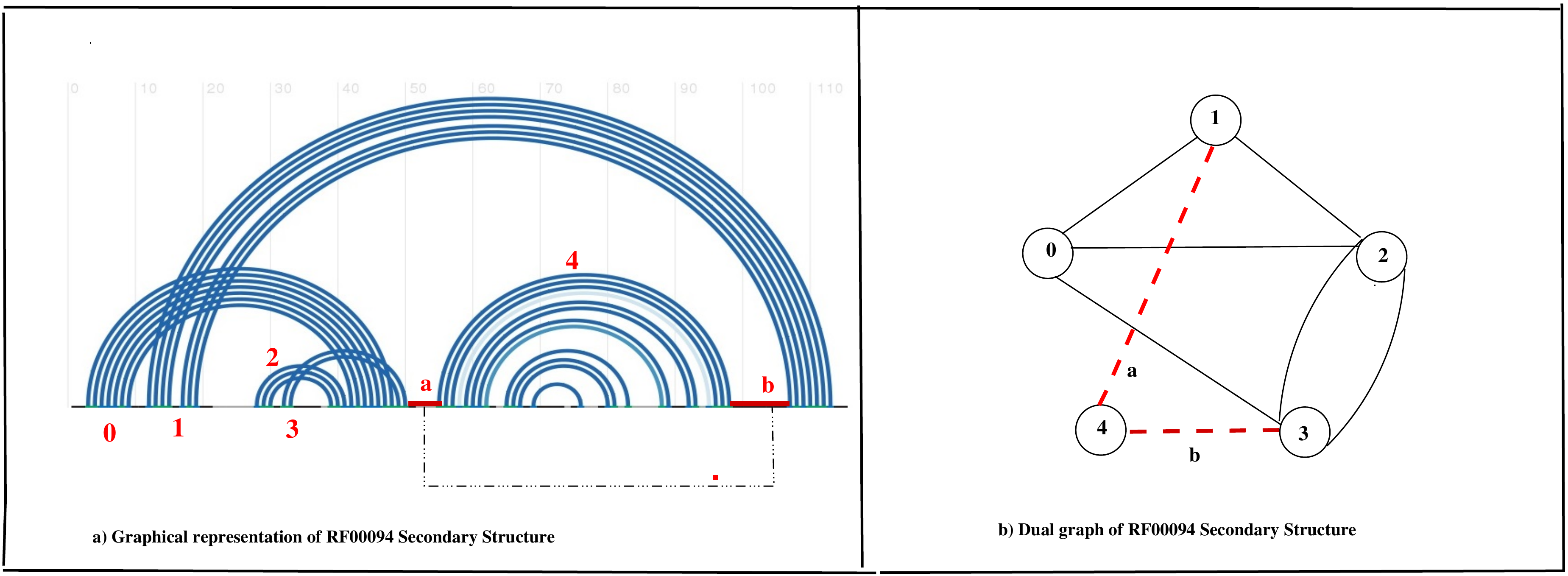}
\end{center}
\caption{\it Hepatitis Delta Virus Ribozyme secondary structure. a) Graphical representation. b) Dual graph representation.} \label{fig:rf00094}
\end{figure}

A pseudoknotted block can be classified as recursive by determining the edge-connectivity of the block. The {\it  edge-connectivity} is defined as the minimum number of edges that if they are deleted then the resulting graph is disconnnected. As an example consider the {\it Hepatitis Delta Virus Ribozyme} (see Fig.~\ref{fig:rf00094}), necessary for viral replication. The stem labeled $4$ in the graphical representation (or vertex labeled $4$ in the dual graph) is attached to the pseudoknot by the segments $a$ and $b$ in its graphical representation, or edges labeled $a$ and $b$ in the dual graph representation. As by deleting two edges in the dual graph, vertex labeled $4$ becomes disconnected, then stem $4$ is a recursive region of the PK, thus the dua graph edge-connectivity is $2$. The proof of the following lemma was shown in \cite{PetSch2}.

\begin{lemma}\label{PKconn} The dual graph representation of a pseudoknotted  block is recursive if and only if the block has edge-connectivity 2.
\end{lemma}

The edge-connectivity of a graph $G=(V,E)$ can be determined in polynomial time in order $ (|V| |E|^2)$ using the max-flow min-cut theorem of network flows by Edmond and Karp~\cite{EK72}, or Ford and Fulkerson \cite {FF56}. 

We can also delete each pair of edges and determine if the graph is disconnected using Depth-First-Search \cite{harary} in time $(|E|^3)$; this variation allows us to find every internal recursive region of the recursive pseudoknot if such pair of edges exist.

The following is the partitioning and classification of pseudoknots algorithm. 
\setcounter{theorem}{1}
\begin {algorithm}\label{algo:recursive}
{\bf Partitioning and Classification of PKs}
\begin {itemize}
\item [i.] Input dual graph $G=(V,E)$ as the Adjacency Matrix, of a  RNA 2D. 
\item [ii.] Output partitioning of the RNA 2D into recursive PK, non-recursive PK, and regular regions.
\begin {enumerate}
      \item {\em Partition the dual graph into blocks by application of Hopcroft and Tarjan's algorithm;}
      \item {\em Analyze each block to determine whether each contains a vertex of degree at least 3}; 
      \item {\bf IF } the block has a vertex of degree $\ge 3$ {\it then} the block is a {\bf pseudoknot;} \\
                              \begin {itemize} 
                                   \item  Apply max-flow min-cut theorem to determine edge-connectivity;
                                   \item  {\it if } edge-connectivity = 2 then the block is a {\bf recursive pseudoknot;} \\
                                              {\it  else} the {\bf pseudoknot is not recursive;} \\
                              \end {itemize}
   \item {\bf ELSE}  the {\bf block is a regular region;}.
  \end {enumerate}
   \end {itemize}
\end {algorithm}  
\setcounter{theorem}{3}

   

\begin{figure}[t]
\begin{center}
\includegraphics [scale=0.43]{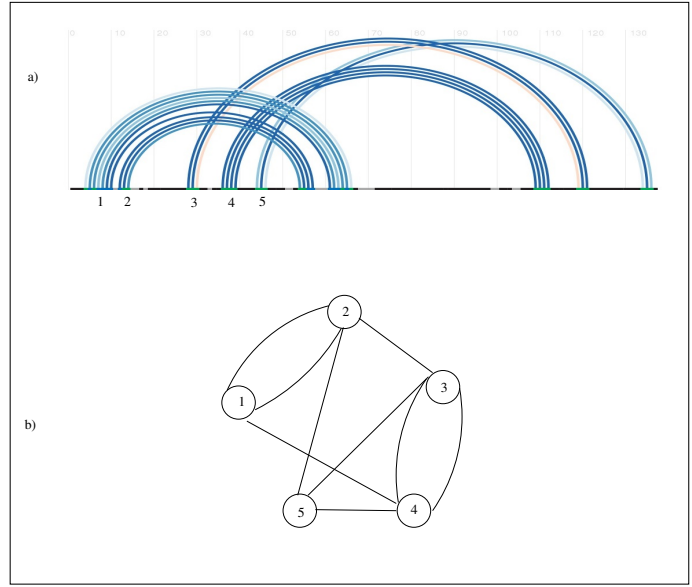}
\end{center}
\caption{{\it Translational repression of the Escherichia coli alpha operon mRNA}. a) Graphical representation; b) Dual graph representation.} \label{fig:PKB236}
\end{figure}

As an example of a non-recursive pseudoknot consider the {\it Translational repression of the Escherichia coli alpha operon mRNA} (\cite {Sch2001}), illustrated in Fig.~\ref{fig:PKB236}. The dual graph representation of this motif 2D has edge-connectivity 3, thus it is not a recursive PK.

 The Appendix illustrates the output generated by the modified algorithm when is run on some of the aforementioned motifs. The algorithm is written in C++ and is archived for public use \cite{PetCode}.





\section {Classification of H, K, L, M pseudoknots types}
\label{S3}

In the classical literature, pseudoknots have been classified and predicted by folding algorithms into four types: H,K,L, and  M [22].  Even though each type is defined in terms of how few stems intertwined, pseudoknots can be complex structures, recursive, and be comprised of several stems. In this paper we show based on dual directed graphs, each PK type can be identified through a series of reductions to an unique representative. This methodology will allow us to systematically analyze thousands of motifs and develop more precise RNA folding algorithms. The results and algorithms discussed in Section \ref{S2.1} and Section \ref{S2.2}, based upon undirected dual graphs, can be easily extended to dual directed graphs, as the direction of a directed edge can be ignored.

In this section we follow the definitions of pseudoknots as by the papers of  Kucharik et al [22] and Antczak et al. [1]. 

\begin{figure}[t]
\begin{center}
\includegraphics[scale=.34]{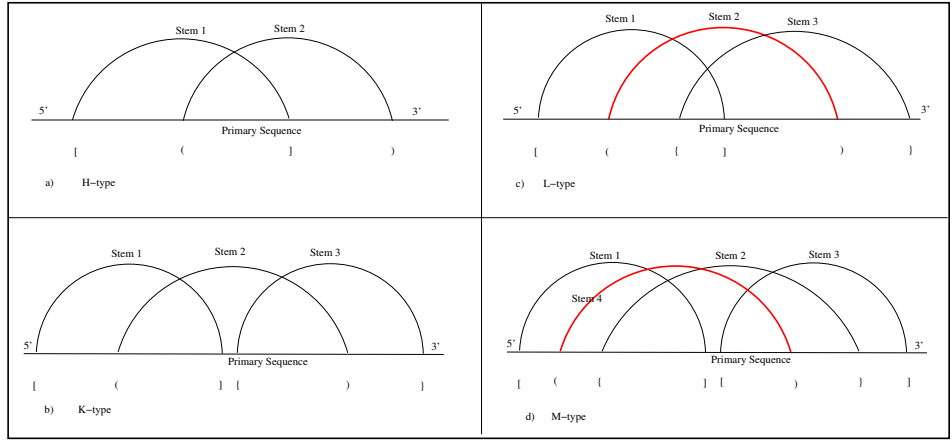}
\end{center}
\caption{{\it Graphical representation of the H,K,L, and M types.  a) H-type. b) K-type, c) L-type, and d) M-type.}} \label{fig:M-K-L-M}
\end{figure}

A H-type pseudoknot occurs when a nucleotide of a loop or bulge pairs with a nucleotide of a single-stranded region outside the loop; this type can be alternatively illustrated  from the graphical representation (see Fig.~\ref{fig:M-K-L-M}-a) by the intertwining of two stems. 

A K-type PK results when two nucleotides from different loops (or bulges) pair to form a double helical segment  (see Fig.~\ref{fig:M-K-L-M}-b). Similarly we can also describe L and M types pseudoknots,  derived from the H and K types, respectively, by addition of a stem (colored red) as shown in Figure~\ref{fig:M-K-L-M}-c and  Figure~\ref{fig:M-K-L-M}-d. 

\begin{figure}[t]
\begin{center}
\includegraphics[scale=.65]{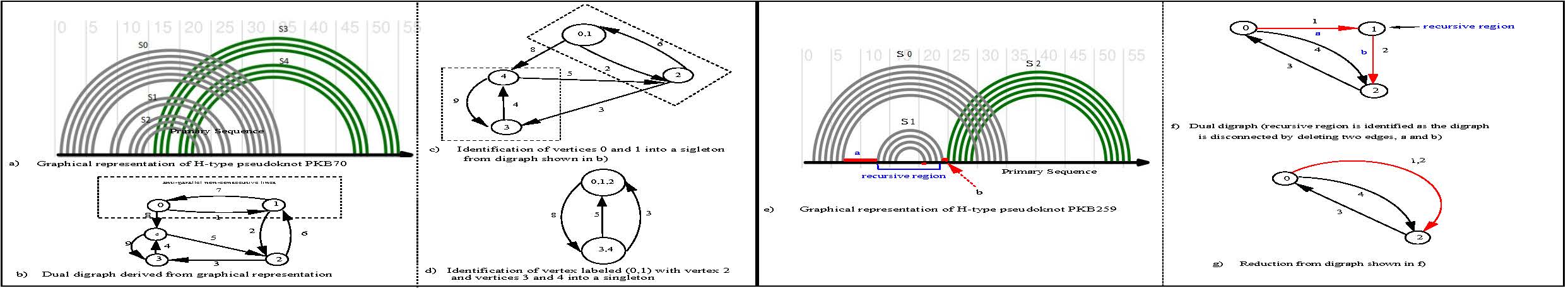}
\end{center}
\caption{{\it Reduction of H-type pseudoknots for RNAs PKB70 and PKB259 to the dual digraph representative of a H-type PK.}} \label{pkb70-259-2}
\end{figure}

Even though the different PK types are defined in terms of few stems, each type may be composed of several stems and also they could be recursive (see section~\ref{S2.2}). For example Figure~\ref{pkb70-259-2} illustrates two H-type PKs, PKB70 (i.e., PK PK4 of Legionella pneumophilia tmRNA), causative agent of the Legionannary's disease [21], composed of 5 stems, and recursive PKB259 (i.e., potato yellow vein virus [27]) comprising a recursive region and 3 stems.
Even though two PKs of the same type of maybe structurally different, in the following section we show how using dual digraphs each motif type can be reduced to a unique digraph type-representative. 

\begin{figure}[bth]
\begin{center}
\includegraphics[scale=.28]{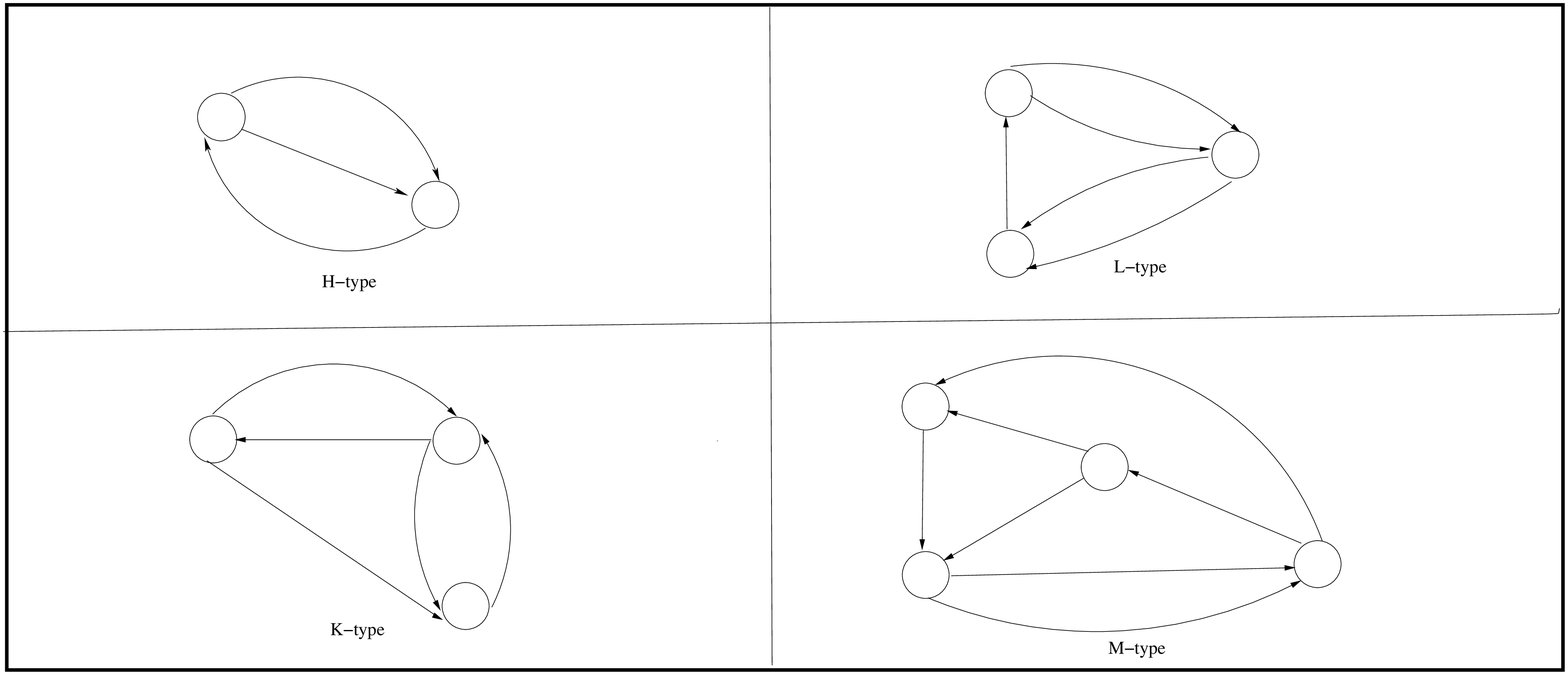}
\end{center}
\caption{{\it By application of Reductions 1 and 2, a directed graph representing a RNA motif,  will be reduced to a unique representative of the 4 different types (H,K,L,M).}} \label{fig:types}
\end{figure}

\subsection {Dual digraphs and reductions to unique PK type identifiers}
\label{3.1}

In this section we used dual digraphs and based upon two graph-theoretical transformations, these digraphs can be reduced to a unique representative identifying  H,K,L, and M pseudoknot types.

A dual digraph can be easily derived from the graphical representation of a RNA 2D structure: each stem is modeled by a vertex of the dual digraph, and following the primary sequence in linear order (i.e., from the 5' end to the 3' end), a segment between stems $S_i$ and $S_j$ is represented by an directed edge $(S_i, S_j)$ in the dual digraph. Moreover each directed edge is assigned a weight corresponding to the order in which the segment is reached following the primary sequence (see Fig.~\ref{pkb70-259-2}-b); this weight is represented by $w(S_i,S_j)$.

From a RNA 2D perspective, we informally define  a {\it super-stem} as a stem that completely contains another stem ({\it sub-stem}). Formally a super-stem is a $n$-tuple  $(x_i, x_{i+1},$
$\ldots,$ 
$x_{i+r}, x_{j-r}\ldots, x_{j-1}, x_j)$, $n=j-i + 1$; in which $(x_{i+k},x_{j-k}), k \leq r$ forms a base-pair (see definition \ref{def1}-d), if there exist another stem (i.e., sub-stem) in the region $(x_l, x_{l+1},\ldots,x_{l+s}, $ 
$x_{m-s},\ldots, x_{m-1}, x_m)$,  $l > i+r$, and $m <  j-r$. For example in Stem $S_0$ shown is Figure~\ref{pkb70-259-2}-a (i.e., PKB70) is a super-stem of $S_1$ and $S_2$, and stem $S_1$ is a super-stem of $S_2$; similarly stem $S_3$ is a super-stem with respect to $S_4$.  For ease of notation let $S_i > S_2$ represent the case when  $S_i$ is a super-stem of $S_j$.
The following proposition describes how a super-stem $S_i$ and its corresponding sub-stem $S_j$ can be detected in a dual digraph,

\begin{proposition}\label{PKred-1} If there are exist exactly two anti-parallel directed edges $(S_i,S_j)$ and $(S_j,S_i)$ between vertices $S_i$ and $S_j$  of the dual digraph, with $w(S_j,S_i) - w(S_i,S_j) \geq 2$, then  $S_i$ is a super-stem of $S_j$.
\end{proposition}

\begin {proof} Without loss of generality, let $w(S_j,S_i) >  w(S_i,S_j)$. Since $w(S_j,S_i) - w(S_i,S_j) \geq 2$, implies that if there is either an eternal stem $S_k$, intertwining $S_i$ {\bf and}  $S_j$, or, $S_j$ forms a self-loop; if there exist an external stem $S_k$ intertwining $S_i$ and $S_j$, this stem does not intercept the primary sequence  between $S_i$ and $S_j$. If $w(S_j,S_i) - w(S_i,S_j) = 1$, then $S_j$ is not fully contained in $S_i$ (i.e., intertwines $S_j$).

\end {proof}

As an example consider stems labeled $S_0$ and $S_1$ for PKB270 shown in Figure~\ref{pkb70-259-2}-a. In the corresponding dual digraph (Fig.~\ref{pkb70-259-2}-b) the vertices labeled $S_0$ and $S_1$ are connected by exactly two anti-parallel edges with $w(S_1,S_0) - w(S_0,S_1) = 6 \geq 2$.
Please note that the converse it is not necessarily true, that is, if $S_i$ is a super-stem of $S_j$, then there exist exactly two anti-parallel edges between vertices $S_i$ and $S_j$ in the dual digraph with  $w(S_1,S_0) - w(S_0,S_1) \geq 2$; for example in PKB259 (Fig.~\ref{pkb70-259-2}-e), stem $S_0$ is a super-stem of $S_1$, however there are no anti-parallel edges between the corresponding vertices in the dual digraph (Fig.~\ref{pkb70-259-2}-f). This has to do with the fact that another stem (i.e., $S_3$)  intercepts the primary sequence between $S_0$ and $S_1$.
From Proposition~\ref{PKred-1} we introduce the following graph reduction,
\setcounter{theorem}{0}
\begin{reduct} \label {One}  Let $G=(V,E)$ be a dual digraph.If there are exist exactly two anti-parallel directed edges $(S_i,S_j)$ and $(S_j,S_i)$ between vertices $S_i$ and $S_j$  of $G$, with $w(S_j,S_i) - w(S_i,S_j) \geq 2$, then transform $G$ into $G'$, where  $S_1$ and $S_j$ are identified into a single vertex $[S_i,S_j]$ and any directed edge in $G$, $(S_k,S_i)$, $(S_k,S_j)$, $k \neq i$ or $k \neq j$, will  result in an directed edge $(S_k,[S1,S2])$ in $G'$, while keeping the same weights from the edges of $G$. Similarly an edge $(S_i,S_k)$, or $(S_j,S_k)$  in $G$ will have the corresponding edge $([S1,S2], S_k)$ in $G'$, while maintaining the same weights on the original edges (see Fig.~\ref{pkb70-259-2}-b-c).\end {reduct}

A recursive PK can be classified  by our partitioning algorithm and a recursive fragment can be identified (see Section~\ref{S2.1}). For example in the graphical representation of PKB259 (Fig.~\ref{pkb70-259-2}-e), segments $a$ and $b$ isolate the recursive fragment Stem $S_1$, or equivalently, deleting exactly two edges of the dual digraph (connectivity 2), $a$ and $b$, identified by Algorithm \ref{algo:recursive}, disconnects the dual digraph. The following graph reduction follows from the algorithm,

\begin{reduct} \label {Two} Let $G=(V,E)$ be a dual digraph, and  $G_0$ be a recursive fragment adjacent to Stem $S_0$ by direct segment $a$ from Stem $S_0$ to $G_0$, and by segment $b$ from $G_0$ to Stem $S_1$  (i.e., deleting segments $a$ and $b$ isolates $G_0$), then transform $G$ into $G'$, by deleting segments $a$ and $b$, recursive fragment $G_0$, and by inserting a direct segment from $S_0$ to $S_1$ (see Fig.~\ref{pkb70-259-2}-e-f).\end {reduct}

By application of Reductions \ref {One} and \ref {Two}, a directed weighted dual graph modeling a pseudnotted region of a RNA 2D (recognized by partitioning Algorithm \ref{algo:recursive}, Section \ref{S2.2}) will be reduced to one types depicted in Fig \ref{fig:types}, otherwise the PK is none of these types.

\section {Conclusions and Ongoing Work}
\label{S4}
The Covid-19 pandemic accelerated the study of viral replication and the need to develop therapeutics to control infectivity. For example, a single mutation of a single nucleotide in the frameshifting region of the SARS-CoV-2 RNA results in a concomitant obliteration of viral replication. The importance of the three-stemmed pseudoknot-dependent ribosomal frameshifting for the propagation of SARS-related coronaviruses it is well-established. We suggest that pseudoknots not only play a significant role, but a predominant one in viral transmissibility for most viruses, and our proposed techniques aim to shed some light in this area, as well to better understand the roles of PKs in general RNA functionality.

\section {Appendix}
\begin{figure} [t]
\begin{center}
\includegraphics[scale=.3]{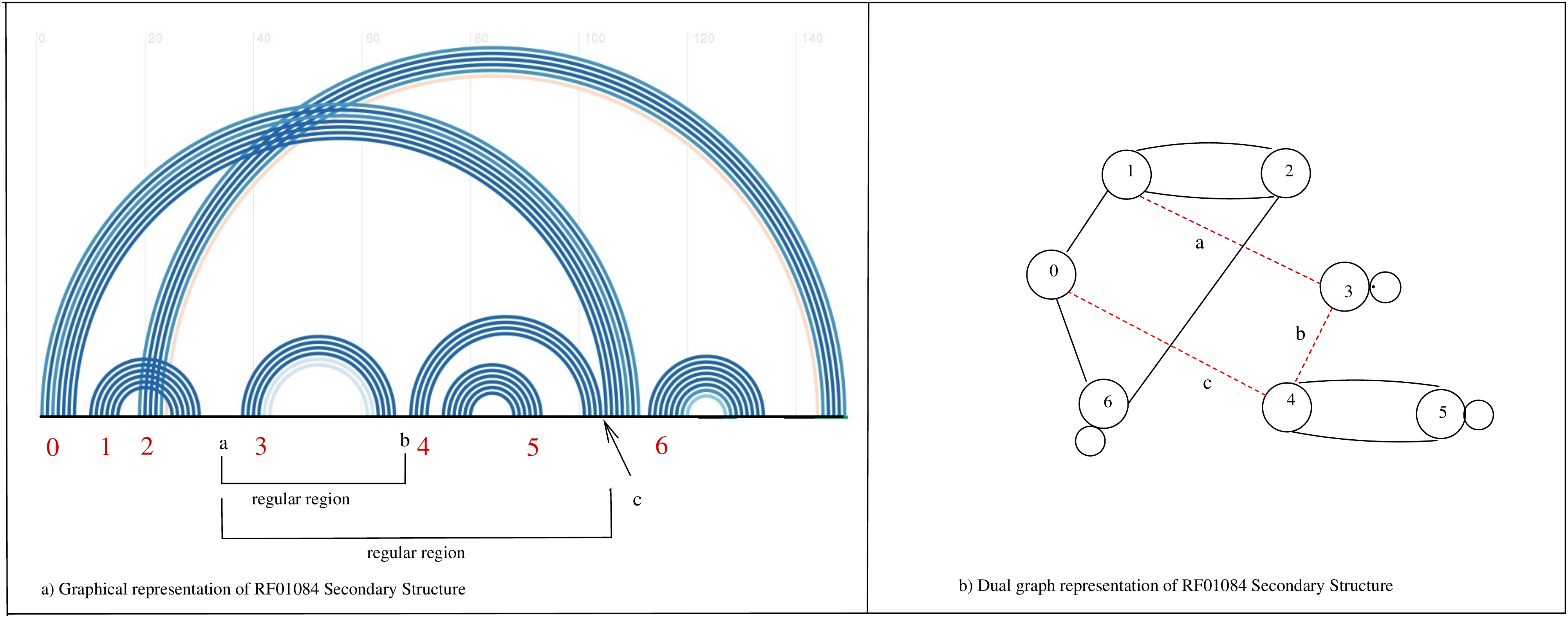}
\end{center}
\caption{ \it A tRNA-like-structure. a) Graphical representation. b) Dual graph representation.} \label{fig:rf01084}
\end{figure} 

Let $(a,b)$ represents an edge of a dual graph with end-vertices $a$ and $b$.




We are next  illustrating the output of the partitioning algorithm tested on the tRNA-like-structure dual graph (see Fig. \ref{fig:rf01084}). 

--------------------- Motif :RF01084 -----------------------------\\
===================== New Block ======= \\
(4,5) - (4,5) - \\
---- this block represents a regular-region ----\\ 
===================== New Block ======= \\
(4,0) - (3,4) - (1,3) - (6,0) - (2,6) - (1,2) - (1,2) - (0,1) - \\
removed edges (2,6) and (6,0), these two edges are a disconnecting set:\\
The block is a recursive PK.\\
----------- Summary information for Motif :RF01084 --------------------------------\\
----------- Total number of blocks: 2\\
----------- number of non-recursive PK blocks: 0\\
----------- number of recursive PK blocks: 1\\
----------- number of regular blocks : 1\\
------------------------------------------------------------------------------------
 
 \section *{Acknowledgments} The work of L. P. was supported by 
PSC-CUNY Award $\#$ 61249-00-49 of the City University of New York. \\

\end{document}